\documentstyle[12pt,mssymb,equations]{article}
\newcounter{eqnpart}  
\renewcommand{\root}{\mbox{$\frac{1}{\sqrt{2}}$}}
\renewcommand{\cite}[1]{\mbox{$^{#1}$}} 

\renewcommand{\theequation}{\arabic{section}.\arabic{equation}\roman{eqnpart}}

\newcommand{\striated}[1]{{\Bbb #1}}
\newcommand{\bold}[1]{\mbox{\boldmath $#1$}}
\newcommand{\lie}[1]{\raisebox{-1.0ex}   
  {$\stackrel{\textstyle {\cal L}}{\bold{\scriptstyle #1}}$}}
\newcommand{\tensor}[2]  
  {(\raisebox{-.6ex}{$\stackrel{#1}{\scriptstyle #2}$})-tensor}
\newcommand{\pseudotensor}[2]  
  {(\raisebox{-.6ex}{$\stackrel{#1}{\scriptstyle #2}$})-pseudotensor}
\newcommand{\cbar}{\mbox{$\bar{\mbox{c}}$}}
\newcommand{\stepup}{\; \raisebox{-.3ex}{$\ulcorner$} \;}   
\newcommand{\stepdown}{\; \raisebox{-.3ex}{$\urcorner$} \;} 

\newcommand{\C}{\striated{C}} 

\newcommand{\PCP}{\striated{P}\striated{C}\striated{P}}
\renewcommand{\C}{{\cal C}} 

\newcommand{\gurses}{G\"{u}rses}
\begin{document}

\title{Generalized Symmetries of the Einstein Equations}
\author{Frederick J.\ Ernst and Isidore Hauser\thanks{Home address:
	4500 19th Street, No.\ 342, Boulder, CO 80304.} \\
	FJE Enterprises\thanks{A private company supporting
	research of F.\ J.\ Ernst and I.\ Hauser
	in classical general relativity theory.} \\
	Rt.\ 1, Box 246A, Potsdam, NY 13676}
\maketitle
\begin{abstract}
We reformulate the symmetries of \gurses\ [Phys.\ Rev.\ Lett.\ {\bf 70},
367 (1993)] in a more abstract, more geometrical manner.  The type (b)
transformation of \gurses\ is related to a diffeomorphism of the
differentiable manifold onto itself.  The type (c) symmetry is replaced by
a more general type (\cbar) symmetry that has the nice property that the
commutator of a type (\cbar) generator with a type (a) generator is itself
of type (\cbar).  We identify a differential constraint that transformations
of type (c) and (\cbar) must satisfy, and which, in our opinion, may
severely limit the usefulness of these transformations.
\end{abstract}
\vfill

\newpage
\section{Introduction}
Not only do the Einstein equations describe the curvature of
spacetime produced by material sources, including curvature variations
associated with gravitational radiation, they also determine the motion
of those material sources.  It would seem, therefore, too much to
expect that methods that have proved so successful for solving Einstein's
equations in the presence of isometries (e.g., two commuting Killing
vectors) would ever be extended to the full theory.  Nevertheless, if no
one is bold enough to look for such extensions, they will certainly never
be found.  Therefore, we were delighted when we heard about the
discovery\cite{1} of new generalized symmetries of the vacuum field
equations, and we immediately set about trying to see if anything could
be achieved by exploiting these symmetries.

The first question was ``Are these transformations merely gauge?''  If
they are not, then can one construct an example that illustrates the
alteration of a spacetime to which such a transformation is applied?
Indeed, quite early in our investigations we thought we had just such an
example, involving a \gurses\ type (b) transformation on a Petrov type
N spacetime.  We subsequently found a flaw in our type N argument, but
\gurses\ himself showed that a similar argument does work in the case of
his type (c) transformation.\cite{2}  One {\em can} produce a general
pp-wave from a mere plane-wave.

In the course of our study we developed a more abstract, more geometrical
formulation of the symmetries of the vacuum Einstein equations.  The
type (b) symmetry we described in terms of a real vector field, and
we replaced \gurses' type (c) symmetry by a more general symmetry we call
(\cbar), involving an antihermitian traceless tensor field.  The (\cbar)
symmetry has the nice property that the commutator of a type (\cbar)
generator and a type (a) generator is itself of type (\cbar), while the
commutator of a type (c) generator and a type (a) generator is {\em not}
of type (c).  Unfortunately, we also found that the type (\cbar) generator
is subject to a nasty differential constraint.  This constraint, which
applies also to the type (c) generator, is satisfied in the case of the
pp-waves application we mentioned earlier, but it is by no means clear
how to cope with this constraint more generally.

In this paper, we shall relate the type (b) symmetry to diffeomorphisms
of the differentiable manifold, and we shall identify the constraints that
restrict the type (c) and type (\cbar) generators.  In this way, we hope
to resolve an apparent conflict between results announced by
\gurses\ and by C.\ Torre and I.\ Anderson.\cite{3}

\section{The Linearized Vacuum Field Equations}
Generally we have found it advantageous in our work to employ geometric
objects, such as the curvature \tensor{2}{2}\ $\striated{R}$ and that
\pseudotensor{2}{2}\ $\striated{D}$ which is the duality operator for
$2$-forms, directly (as opposed to using their components relative to
a coordinate tetrad).  The tensors $\striated{R}$ and $\striated{D}$
are given by Eqs.\ (\ref{Rdef}) and (\ref{Ddef}) in the appendix to this
paper.  The projection operator $\striated{P}$ given by Eq.\ (\ref{Pdef}),
and its complex conjugate $\striated{P}^{*}$, divide the six-dimensional
space of $2$-forms into three-dimensional self-dual (eigenvalue $+i$) and
anti-self-dual (eigenvalue $-i$) parts.  If $\{ k,m,t,t^{*} \}$ is any
given tetrad of null $1$-forms, a convenient basis for self-dual
$2$-forms is given by $B_{1},B_{0},B_{-1}$ in Eq.\ (\ref{Bdef}).

We denote the corresponding members of the tangent space by bold type.
Thus, if $e^{b}$ is any orthonormal tetrad of $1$-forms, the corresponding
orthonormal tetrad of $1$-vector fields will be denoted by $\bold{e}_{a}$,
where $\bold{e}_{a}e^{b}=\delta_{a}^{b}$.  Observe that when we regard a
differential form such as $e^{b}$ as a linear functional on the tangent
space, we write it to the right of the member of the tangent space upon
which it operates, while if we write a $p$-form to the left of a
$q$-vector field, we mean a \tensor{p}{q}.  Generally speaking, we employ
striated symbols such as $\striated{R}$ and $\striated{D}$ for such
objects.  We customarily suppress the symbol $\wedge$ in exterior products
of differential forms, although we retain it in the exterior products of
vector fields.

For the purposes of this paper, we shall describe the symmetries of the
vacuum Einstein equations in terms of a \tensor{1}{1}\ $\striated{L}$ and
a \pseudotensor{2}{1}\ $\striated{M}$, which are
defined by
\begin{subequations}
\begin{eqnarray}
\striated{L} & := & (\delta e^{b}) \bold{e}_{b} \; , \label{Ldef} \\
\striated{M} & := & \striated{N} \stepdown \striated{I} \; ,
\label{Mdef}
\end{eqnarray}
\end{subequations}
where
\begin{equation}
\striated{N} := (\delta v) \bold{B}_{-1}
- \frac{1}{2} (\delta u) \bold{B}_{0}
+ (\delta w) \bold{B}_{1}
\label{Ndef}
\end{equation}
and $\delta u, \delta v, \delta w$ are the variations of the complex
connection $1$-forms defined in Eqs.\ (\ref{uvwdef}), $\striated{I}$ is
the unit \tensor{1}{1}\ defined in Eq.\ (\ref{Idef}), and $\stepdown$
is the Grassmann inner product for the tangent manifold, which has the
properties (\ref{grass1}) and (\ref{grass2}).  There is a corresponding
Grassmann inner product $\stepup$ for the cotangent manifold that has
the analogous properties (\ref{grass3}) and (\ref{grass4}).

In terms of these invariant objects the linearized vacuum field equations
assume the simple forms
\begin{subequations}
\begin{eqnarray}
d\striated{L} + \striated{M} + \striated{M}^{*} & = & 0 \; ,
\label{eq1} \\
d\striated{M} + (\PCP) \stepdown \striated{L} & = & 0 \; ,
\label{eq2}
\end{eqnarray}
\end{subequations}
where $\striated{C}$ is the Weyl conform tensor (\ref{Cdef}), and we
note that
\begin{equation}
\striated{C} \striated{P} = \striated{P} \striated{C} = \PCP \; .
\end{equation}
Not just any solution of Eqs.\ (\ref{eq1}) and (\ref{eq2}) will do!
Because of the definitions (\ref{Mdef}) and (\ref{Ndef}), one must
select $\striated{M}$ in such a way that
$\striated{M} = \striated{N} \stepdown \striated{I}$, where
\begin{equation}
\striated{N} \striated{P}^{*} = 0 \; .
\label{essential}
\end{equation}

We have the following important uniqueness theorem.
\begin{description}
\item[Theorem 1:]
If, for a given initial solution of the vacuum field equations, both
$(\striated{L}, \striated{N})$ and $(\striated{L}, \striated{N}')$
satisfy the linearized vacuum field equations, then $\striated{N}' =
\striated{N}$.
\end{description}
{}From Eq.\ (\ref{eq1}), one has
\begin{equation}
\striated{M}' + \striated{M}^{\prime *} =
\striated{M} + \striated{M}^{*} \; .
\end{equation}
{}From the above equation and the relation $\striated{M} := \striated{N}
\stepdown \striated{I}$, it follows that
\begin{equation}
\striated{N}' + \striated{N}^{\prime *} =
\striated{N} + \striated{N}^{*} \; .
\label{real}
\end{equation}
However, $\striated{N}$ and $\striated{N}'$ are \tensor{2}{1}s that
satisfy (\ref{essential}).  This fact may be used, together with
Eq.\ (\ref{real}) to show that $\striated{N}' = \striated{N}$.

Next, suppose that $(\striated{L}^{(1)}, \striated{N}^{(1)})$ and
$(\striated{L}^{(2)}, \striated{N}^{(2)})$ are any solutions of the
linearized vacuum field equations corresponding to the same initial
solution of the vacuum field equations.  Then, it is easily verified
that
\begin{subequations}
\begin{eqnarray}
[\delta^{(1)},\delta^{(2)}] e_{a} & = & \striated{L}^{(1,2)} e_{a} \; , \\
\striated{L}^{(1,2)} & = & \delta^{(1)} \striated{L}^{(2)}
- \delta^{(2)} \striated{L}^{(1)}
+ [\striated{L}^{(2)},\striated{L}^{(1)}] \; .
\label{commutator}
\end{eqnarray}
\end{subequations}
Now, let $\bold{e}$ denote the column matrix whose elements are
$\bold{e}_{a} (a=1,\cdots,4)$.  Then, we can write
\begin{equation}
\striated{L}^{(i)} = e^{T} L^{(i)} \bold{e} \; ,
\end{equation}
where $L^{(i)}$ is a $4 \times 4$ matrix whose elements are real fields.
It follows that
\begin{eqnarray}
\striated{L}^{(1,2)} & = &
e^{T} [\delta^{(1)} L^{(2)} - \delta^{(2)} L^{(1)}] \bold{e}
+ (\delta^{(1)} e^{T}) L^{(2)} \bold{e}
+ e^{T} L^{(2)} (\delta^{(1)} \bold{e}) \nonumber \\ & &
- (\delta^{(2)} e^{T}) L^{(1)} \bold{e}
- e^{T} L^{(1)} (\delta^{(2)} \bold{e})
+ [\striated{L}^{(2)},\striated{L}^{(1)}]
\nonumber \\ & = &
e^{T} [\delta^{(1)} L^{(2)} - \delta^{(2)} L^{(1)}] \bold{e}
+ [\striated{L}^{(1)},\striated{L}^{(2)}] \; .
\end{eqnarray}

\subsection{Antisymmetric Case:}
When $\striated{L}$ is antisymmetric, we can introduce a $2$-vector field
\begin{equation}
\bold{L} := \frac{1}{2} \striated{I} \stepup \striated{L} \; .
\end{equation}
Then
\begin{subequations}
\begin{eqnarray}
\striated{L} & = & \striated{K} + \striated{K}^{*} \; , \\
\striated{K} & := & (\bold{L} \striated{P}) \stepdown \striated{I} \; .
\end{eqnarray}
\end{subequations}
Suppose further that
\begin{subequations}
\begin{eqnarray}
\striated{N} & = & -d(\bold{L}\striated{P}) \; , \\
\striated{M} & = & -d\striated{K} \; ,
\end{eqnarray}
\end{subequations}
which satisfies the essential condition (\ref{essential}) because
$d\striated{P}^{*}=0$.  Then Eq.\ (\ref{eq1}) will obviously be
satisfied, while
\begin{eqnarray}
d\striated{M} + (\PCP) \stepdown \striated{L} & = &
- \striated{C} \stepdown \striated{K}
+ (\PCP) \stepdown (\striated{K} + \striated{K}^{*}) \nonumber \\
& = & - (\PCP)^{*} \stepdown \striated{K}
+ (\PCP) \stepdown \striated{K}^{*} \; .
\label{second}
\end{eqnarray}
At this point we use the fact that $\bold{L} \striated{P}$ is a
self-dual $2$-vector field.  For any \tensor{p}{2}\ $\striated{U}$
such that $\striated{U} \striated{P}^{*} = 0$, one has
\begin{equation}
(\PCP)^{*} \stepdown (\striated{U} \stepdown \striated{I}) = 0 \; .
\end{equation}
Hence each of the terms on the right side of Eq.\ (\ref{second})
separately vanishes, and the second linearized vacuum field equation
(\ref{eq2}) is satisfied.  Thus we have the first symmetry.
\begin{description}
\item[Definition:]
A type (a) transformation is one for which $\striated{L}$ is antisymmetric,
while
\begin{subequations}
\begin{eqnarray}
\striated{M} & = & d(\bold{L}\striated{P}) \stepdown \striated{I} \; ,
\label{sM} \\
\bold{L} & := & \frac{1}{2} \striated{I} \stepup \striated{L} \; .
\label{sL}
\end{eqnarray}
\end{subequations}
\end{description}

When $\delta L^{ab}=0$, the type (a) transformation is readily
exponentiated to give the usual Lorentz transformation field of an
orthonormal tetrad.  When $\delta L^{ab} \ne 0$, the type (a)
transformation is still an infinitesimal Lorentz transformation,
and is thus purely gauge, since for the variation of the abstract
metric tensor one obtains
\begin{equation}
\delta g = \delta (e^{a} \otimes e_{a})
= (\delta e^{a}) \otimes e_{a} + e^{a} \otimes (\delta e_{a}) = 0 \; .
\end{equation}

\subsection{Some Simple Generalized Symmetries}
Another solution of the linearized vacuum field equations (\ref{eq1})
and (\ref{eq2}) and the essential condition (\ref{essential}) is the
following one.
\begin{description}
\item[Definition:]
A type (b) transformation is one for which
\begin{subequations}
\begin{eqnarray}
\striated{L} & = & - d\bold{a} \; , \\
\striated{M} & = & (\PCP) \stepdown \bold{a} \; ,
\end{eqnarray}
\end{subequations}
\end{description}
where $\bold{a}$ is an arbitrary real vector field and $a$ is the $1$-form
that is the coform of $\bold{a}$.
To show this is a solution of Eqs.\ (\ref{eq1}) and (\ref{eq2}) use the
relation (\ref{d2}) together with the Bianchi integrability condition
(\ref{Bianchi}).  That the essential condition (\ref{essential}) is
satisfied follows from the fact that the selected $\striated{M}$ is
equivalent to the selection
\begin{equation}
\striated{N} = a \stepup (\PCP) \; .
\end{equation}
This can be shown using the algebraic identify (\ref{algebraic}).  In the
next section we shall show that this transformation is identical to
\gurses' type (b) transformation, and later we shall relate this symmetry
to spacetime diffeomorphisms.

\begin{description}
\item[Definition:]
The {\em scaling transformation} is defined by
\begin{subequations}
\begin{eqnarray}
\striated{L} & = & \lambda \striated{I} \; , \\
\striated{N} & = & 0 \; , \\
\striated{M} & = & 0 \; ,
\end{eqnarray}
\end{subequations}
where $\lambda$ is a real constant.
\end{description}
The essential condition (\ref{essential}) is satisfied, since
$d\striated{I}=0$ and $(\PCP) \stepdown \striated{I} = 0$.  This
transformation corresponds to a constant rescaling of the spacetime
metric.

We should like to stress the importance of checking that the condition
(\ref{essential}) is satisfied before one claims that one has found a
symmetry of the vacuum field equations; it is not sufficient just to
satisfy the linearized field equations (\ref{eq1}) and (\ref{eq2})!

\subsection{Traceless Symmetric Case:}
We have seen that antisymmetric $\striated{L}$ are pure gauge.  Therefore,
let us consider now a traceless symmetric $\striated{L}$, i.e., a
traceless $\striated{L}$ such that
\begin{equation}
\striated{L} \stepdown \striated{I} = 0 \; .
\label{LstepI}
\end{equation}
A lengthy but straightforward calculation reveals that in this case
\begin{subequations}
\begin{eqnarray}
\striated{N} & = & [\striated{P}(d\striated{L})]^{\sim} \; , \\
\striated{M} & = & - \striated{P}(d\striated{L})
+ \striated{I} \stepup \left\{ [\striated{P}(d\striated{L})]
\stepdown \striated{I} \right\} \; , \label{Mexpr} \\
\striated{P}(d\striated{L}) & = & - \striated{M}
+ \frac{1}{2} \striated{I} \stepup (\striated{M} \stepdown \striated{I})
\; , \label{PdL}
\end{eqnarray}
\end{subequations}
where $[\striated{P} (d\striated{L})]^{\sim}=(d\striated{L})^{\sim}
\striated{P}$ is the cotensor of $\striated{P}(d\striated{L})$.
(For any $p$-form $\alpha$ and $q$-vector $\bold{\beta}$, the {\em
cotensor} of $\alpha \bold{\beta}$ is $\beta \bold{\alpha}$, where
$\bold{\alpha}$ is the covector of $\alpha$ and $\beta$ is the coform
of $\bold{\beta}$.)  The linearized field equation (\ref{eq1}) is
satisfied by Eq.\ (\ref{Mexpr}).  From Eq.\ (\ref{PdL}) it is seen
that the essential condition (\ref{essential}) is equivalent to the
following algebraic condition on $\striated{M}$:
\begin{equation}
\striated{P}^{*} \left[ \striated{M} - \frac{1}{2} \striated{I}
\stepup (\striated{M} \stepdown \striated{I}) \right] = 0 \; .
\label{PstarM}
\end{equation}

Suppose now that
\begin{subequations}
\begin{eqnarray}
\striated{L} & = & \striated{K} + \striated{K}^{*} \; , \\
\label{L0def}
\striated{K} & = & i e_{a} \stepup (\PCP) \stepdown \bold{e}_{b}
T^{ab} \; , \label{Kdef} \\
\striated{W} & := & -d\striated{K} \; , \label{Wdef}
\end{eqnarray}
\end{subequations}
where
\begin{equation}
\striated{T} := e_{a} T^{ab} \bold{e}_{b}
\end{equation}
is a real symmetric tensor field.  {\em Then, with the aid of Eq.\
(\ref{id3}), one can prove that $(\striated{L},\striated{W})$ formally
satisfies the linearized field equations (\ref{eq1}) and (\ref{eq2})
when $\striated{M}$ is replaced by $\striated{W}$ in those equations.
However, $\striated{W}$ must not be confused with $\striated{M}$,
since $\striated{W}$ does not generally satisfy the essential
condition (\ref{PstarM})}.

One way of overcoming the fact that $\striated{W}$ generally
differs from $\striated{M}$ follows a procedure equivalent to
one employed by \gurses.\cite{2}  We find that Eq.\ (\ref{Mexpr})
and (\ref{Wdef}) imply that
\begin{subequations}
\begin{eqnarray}
\striated{M} & = & \striated{W} + \striated{S} \; , \\
\striated{S} & = & (\striated{P}\striated{W}^{*}
- \striated{P}^{*}\striated{W})) - I \stepup
[(\striated{P}\striated{W}^{*}
-\striated{P}^{*}\striated{W}) \stepdown I] \; .
\label{Sdef}
\end{eqnarray}
\end{subequations}
Thus the linearized field equation (\ref{eq2}) will be satisfied by
$(\striated{L},\striated{M})$ if and only if
\begin{equation}
d\striated{S} = 0 .
\label{dS}
\end{equation}
The differential constraints on $\striated{T}$ that are imposed by
the above Eq.\ (\ref{dS}) are of the second order, and are cumbersome
except for very special cases.  A somewhat stronger, but simpler,
condition is provided by the following new theorem.
\begin{description}
\item[Theorem 2:]
A necessary and sufficient condition that $\striated{W} = \striated{M}$
is that $\striated{P}^{*}\striated{W}=0$.  When this condition is
satisfied, $(\striated{L},\striated{M})$ is a solution of the
linearized vacuum field equations.
\end{description}
The differential constraints imposed by $\striated{P}^{*} \striated{W}=0$
are of the first order in $\striated{T}$.  Later, in another paper, we
shall return to these differential constraints on the real symmetric
$\striated{T}$.  However, we should now like to identify our type
(\cbar) transformation.
\begin{description}
\item[Definition:]
A type (\cbar) transformation is one for which
\begin{subequations}
\begin{eqnarray}
\striated{L} & = & \striated{K} + \striated{K}^{*} \; , \\
\striated{K} & = & e_{a} \stepup (\PCP) \stepdown \bold{e}_{b}
T^{ab} \; ,
\end{eqnarray}
\end{subequations}
where
\begin{equation}
\striated{T} := e_{a} T^{ab} \bold{e}_{b}
\end{equation}
is an antihermitian \tensor{1}{1}.
\end{description}
This may be regarded as a somewhat artificial joining of rather
different antisymmetric $\striated{L}$ and traceless symmetric
$\striated{L}$ components, but such a joining is necessary if we are
to encompass \gurses' type (c).  We shall see in the next section that
when $\striated{T} = - i a \bold{a}^{*}$, where $\bold{a}$ is a
complex vector field, and $a$ is the $1$-form that is the coform
of $\bold{a}$, then type (\cbar) reduces to type (c).

\setcounter{equation}{0}
\section{Reduction to \gurses' Form}
In \gurses' $2 \times 2$ matrix formulation,
\begin{subequations}
\begin{eqnarray}
\sigma & := & \left( \begin{array}{cc}
m & t^{*} \\
t & -k
\end{array} \right) \; , \\
\Gamma & := & \left( \begin{array}{cc}
\frac{1}{2} u & - w \\
v & - \frac{1}{2} u
\end{array} \right) \; , \\
\bold{B} & := & \left( \begin{array}{cc}
\frac{1}{2} \bold{B}_{0} & - \bold{B}_{-1} \\
\bold{B}_{1} & - \frac{1}{2} \bold{B}_{0}
\end{array} \right) \; ,
\end{eqnarray}
and
\begin{equation}
\C := (\PCP) \stepdown \bold{B} \; .
\end{equation}
\end{subequations}
The identities
\begin{subequations}
\begin{eqnarray}
\bold{B}_{I} \stepdown \bold{\sigma} & = &
(\bold{B}_{I} \stepdown \bold{B}) \bold{\sigma}
\label{id1} \; , \\
\sigma \stepup B_{I} & = & (B \stepup B_{I}) \sigma \; ,
\label{id2}
\end{eqnarray}
\end{subequations}
are especially useful.  Here $B$ is the matrix of $2$-forms that are
coforms of the elements of the matrix $\bold{B}$.

In the case of type (a) we have
\begin{subequations}
\begin{eqnarray}
\delta \sigma & = & \striated{L} \sigma \nonumber \\
& = & (\striated{K} \sigma)
+ (\striated{K} \sigma)^{\dagger} \; , \\
(\delta \Gamma) \sigma & = & \striated{M} \stepdown \bold{\sigma}
\nonumber \\ & = &
- (d\striated{K}) \stepdown \bold{\sigma}
\nonumber \\ & = &
- [d(\striated{K} \stepdown \bold{\sigma})
+ \Gamma (\striated{K} \stepdown \sigma)
+ (\striated{K} \stepdown \sigma) \Gamma^{\dagger}
\nonumber \\ & = &
- [d(\striated{K} \sigma) + \Gamma (\striated{K} \sigma)
+ (\striated{K} \sigma) \Gamma^{\dagger}]
\nonumber \\ & =: &
- D(\striated{K} \sigma) \; ,
\label{Gamsig}
\end{eqnarray}
\end{subequations}
where $\striated{K} = - (\bold{L} \striated{P}) \stepdown \striated{I}$,
and
\begin{eqnarray}
\striated{K} \sigma & = & [(\bold{L}\striated{P}) \stepdown \bold{\sigma}]
\stepdown \striated{I} \nonumber \\ & = &
[\bold{L} \striated{P} \stepdown \bold{B}] \sigma
\nonumber \\ & = &
X \sigma \; ,
\end{eqnarray}
where
\begin{equation}
X := \bold{L} \striated{P} \stepdown \bold{B} \; .
\end{equation}
This may be verified using the identity (\ref{id2}).  Finally,
Eq.\ (\ref{Gamsig}) can be solved, with the result
\begin{equation}
\delta \Gamma = -DX \; .
\end{equation}

In the case of type (b) we have
\begin{subequations}
\begin{eqnarray}
\delta \sigma & = & - (d\bold{a}) \stepdown \bold{\sigma}
\nonumber \\ & = &
- [d(\bold{a} \stepdown \bold{\sigma}) - \bold{a} \stepdown
d\bold{\sigma}]
\nonumber \\ & = &
- [d(\bold{a} \stepdown \bold{\sigma}) + \Gamma (\bold{a} \stepdown
\bold{\sigma}) + (\bold{a} \stepdown \bold{\sigma}) \Gamma^{\dagger}]
\nonumber \\ & = &
- [dA + \Gamma A + A \Gamma^{\dagger}]
\nonumber \\ & =: &
- DA \; , \\
(\delta \Gamma) \sigma & = & \striated{M} \stepdown \bold{\sigma}
\nonumber \\ & = &
[(\PCP) \stepdown \bold{a}] \stepdown \bold{\sigma}
\nonumber \\ & = &
(\PCP) \stepdown (\bold{a} \wedge \bold{\sigma})
\nonumber \\ & = &
- (\PCP) \stepdown (\bold{\sigma} \wedge \bold{a})
\nonumber \\ & = &
- [(\PCP) \stepdown \bold{\sigma}] \stepdown \bold{a}
\nonumber \\ & = &
(\C \bold{\sigma}) \stepdown \bold{a}
\nonumber \\ & = &
\C A \; ,
\end{eqnarray}
\end{subequations}
where
\begin{equation}
\C := (\PCP) \stepdown \bold{B} \; .
\end{equation}
Using the identity (\ref{id1}) one can show that
\begin{equation}
\C \bold{\sigma} = - (\PCP) \stepdown \bold{\sigma}
\label{scriCdef}
\end{equation}
and
\begin{equation}
\C \sigma = 0 \; .
\label{id}
\end{equation}

In the case of type (\cbar) we have
\begin{subequations}
\begin{eqnarray}
\delta \sigma & = & (\striated{K}\sigma) + (\striated{K}\sigma)^{\dagger}
\; , \\
(\delta \Gamma) \sigma & = & - D(\striated{K} \sigma) + s \; ,
\end{eqnarray}
\end{subequations}
where
\begin{eqnarray}
\striated{K} \sigma & = & T_{ab} [e^{a} \stepup (\PCP) \stepdown
\bold{e}^{b}] \sigma
\nonumber \\ & = &
T_{ab} [e^{a} \stepup (\PCP) \stepdown \bold{e}^{b}] \stepdown
\bold{\sigma}
\nonumber \\ & = &
T_{ab} e^{a} \stepup (\PCP) \stepdown (\bold{e}^{b} \wedge \bold{\sigma})
\nonumber \\ & = &
- T_{ab} e^{a} \stepup (\PCP) \stepdown (\bold{\sigma} \wedge
\bold{e}^{b})
\nonumber \\ & = &
- T_{ab} [e^{a} \stepup (\PCP) \stepdown \bold{\sigma}] \stepdown
\bold{e}^{b}
\nonumber \\ & = &
T_{ab} e^{a} \stepup (\C \bold{\sigma}) \stepdown \bold{e}^{b}
\nonumber \\ & = &
T_{ab} (e^{a} \stepup \C) (\bold{\sigma} \stepdown \bold{e}^{b})
\label{Gam}
\end{eqnarray}
and
\begin{eqnarray}
s & := & \striated{S} \stepdown \bold{\sigma} \; ,
\end{eqnarray}
where we have used Eq.\ (\ref{scriCdef}).

Finally, if the antihermitian \tensor{1}{1}\ $\striated{T}$ is assumed to
have the form
\begin{equation}
\striated{T} = - i a \bold{a}^{*} \; ,
\end{equation}
where $\bold{a}$ is an arbitrary complex vector field, and $a$ is the
$1$-form that is the coform of $\bold{a}$, then Eq.\ (\ref{Gam})
reduces to
\begin{equation}
\striated{K} \sigma = - i (a \stepup \C) (\bold{\sigma} \stepdown
\bold{a})^{\dagger} = - i \Omega^{A} A^{\dagger} \; ,
\end{equation}
where
\begin{subequations}
\begin{eqnarray}
\Omega^{A} & := & a \stepup \C \; , \\
A & := & \bold{\sigma} \stepdown \bold{a} \; .
\end{eqnarray}
\end{subequations}
Note, in addition, that
\begin{equation}
\Omega^{A} \sigma = (a \stepup \C) \sigma
= \C (a \stepup \sigma) - a \stepup(\C \sigma)
= \C A \; ,
\end{equation}
where we have used Eq.\ (\ref{id}).  This is \gurses' original type (c)
symmetry.

\setcounter{equation}{0}
\section{Evaluation of Commutators}
Using Eq.\ (\ref{commutator}) we shall now evaluate the commutator
$[\delta^{(1)},\delta^{(2)}]$ for various choices of $\delta^{(1)}$
and $\delta^{(2)}$.

It is instructive to review from our perspective the evaluation of the
commutator of a type (b) generator and a type (a) generator.  Therefore,
let $\striated{L}^{(1)}=-d\bold{a}$ and let $\striated{L}^{(2)}$ be
antisymmetric.  We then find that
\begin{subequations}
\begin{eqnarray}
\delta^{(1)}\striated{L}^{(2)} & = &
[\striated{L}^{(1)},\striated{L}^{(2)}] \; , \\
\delta^{(2)}\striated{L}^{(1)} & = & - d(\delta^{(2)} \bold{a}) \; .
\end{eqnarray}
\end{subequations}
Hence
\begin{subequations}
\begin{eqnarray}
[\delta^{(1)},\delta^{(2)}] & = & -d\bold{c} \; , \\
\bold{c} & := & - \delta^{(2)}\bold{a} \nonumber \\
& = & - a^{b} \delta^{(2)} \bold{e}_{b} \nonumber \\
& = & - a^{b} L_{bc} \bold{e}^{c} \; .
\end{eqnarray}
\end{subequations}
This corresponds to a type (b) transformation with
\begin{equation}
C := \bold{c} \stepdown \bold{\sigma} = - \root a^{b} L_{bc} \sigma^{c}
= \root a^{b} L_{ab} \sigma^{a} = \root a^{b} (X \sigma_{b} + \sigma_{b}
X^{\dagger}) = X A + A X^{\dagger} \; ,
\end{equation}
in terms of \gurses' $2 \times 2$ matrix formalism.  Here we have used
\begin{subequations}
\begin{eqnarray}
\sigma & = & \root e^{b} \sigma_{a} \; , \\
A & = & \root a^{b} \sigma_{a} \; ,
\end{eqnarray}
\end{subequations}
where
\begin{equation}
\sigma_{1} := \left( \begin{array}{cc}
0 & 1 \\
1 & 0
\end{array} \right) \; , \;
\sigma_{2} := \left( \begin{array}{cc}
0 & -i \\
i & 0
\end{array} \right) \; , \;
\sigma_{3} := \left( \begin{array}{cc}
1 & 0 \\
0 & -1
\end{array} \right) \; , \;
\sigma_{4} := \left( \begin{array}{cc}
1 & 0 \\
0 & 1
\end{array} \right) \; . \;
\end{equation}

Now, let us consider the commutator of a type (\cbar) generator and a type
(a) generator.  This time one finds that
\begin{subequations}
\begin{eqnarray}
\delta^{(1)}\striated{L}^{(2)} & = &
[\striated{L}^{(1)},\striated{L}^{(2)}] \; , \\
\delta^{(2)}\striated{L}^{(1)} & = &
\delta^{(2)} [e_{a} \stepup (\PCP) \stepdown \bold{e}_{b}] T^{ab} \; .
\end{eqnarray}
\end{subequations}
Bearing in mind that neither $\striated{C}$ nor $\striated{P}$ is effected
by the type (a) gauge transformation, and the inner product is unaltered,
one finds that
\begin{eqnarray}
\delta^{(2)}\striated{L}^{(1)} & = &
[(\delta^{(2)} e_{a}) \stepup (\PCP) \stepdown \bold{e}_{b} +
e_{a} \stepup (\PCP) \stepdown (\delta^{(2)} \bold{e}_{b})] T^{ab}
\nonumber \\ & = &
[e_{a} \stepup (\PCP) \stepdown \bold{e}_{b}] T^{\prime ab} \; ,
\end{eqnarray}
where
\begin{equation}
T^{\prime ab} = L^{a}_{\; \; c} T^{cb} + L^{b}_{\; \; c} T^{ac}
\end{equation}
is antihermitian.  Thus, the commutator of a type (\cbar) generator
with a type (a) generator is itself of type (\cbar).

\setcounter{equation}{0}
\section{Analysis of the Type (b) Transformation}
There is a resemblance of the equation
\begin{equation}
\delta e^{b} = - (d\bold{a}) e^{b}
\label{deltaE}
\end{equation}
and the equation
\begin{equation}
\lie{a} e^{b} := - \bold{a} (de^{b}) + d(\bold{a}e^{b})
\label{lie}
\end{equation}
that defines the Lie derivative of the $1$-form $e^{b}$ with respect to
the vector field $\bold{a}$.  In the latter equation it should be
understood that $\bold{a}(de^{b})$ is to be computed by using the
Grassmann contraction product, which has the property that
\begin{equation}
\bold{a} (\lambda \mu) = (\bold{a} \mu) \lambda - (\bold{a} \lambda) \mu
\label{grass}
\end{equation}
for any $1$-forms $\lambda$ and $\mu$.  Adding Eqs. (\ref{deltaE}) and
(\ref{lie}), we obtain
\begin{eqnarray}
\delta e^{b} + \lie{a} e^{b} & = &
-(d\bold{a})e^{b} + d(\bold{a}e^{b}) - \bold{a}(de^{b}) \nonumber \\
& = & (d\bold{e}^{b}) a - \bold{a}(de^{b}) \nonumber \\
& = & (\Gamma^{b}_{\; c} \bold{e}^{c}) a - \bold{a}(\Gamma^{b}_{\; c}
e^{c}) \nonumber \\
& = & (\bold{a}\Gamma^{b}_{\; c}) e^{c} \; ,
\label{result}
\end{eqnarray}
where
\begin{equation}
\Gamma_{bc} := (d\bold{e}_{b}) e_{c}
\end{equation}
is the antisymmetric Ricci rotation matrix of connection $1$-forms.
Equation (\ref{result}) implies that the type (b) variation of any given
orthonormal tetrad of $1$-forms $e^{b}$ is the difference of a \gurses\
type (a) variation of $e^{b}$ (that is, an infinitesimal Lorentz
transformation of $e^{b}$) and of the Lie derivative of $e^{b}$ with
respect to $\bold{a}$.

The restriction of the abstract metric tensor $g$ to the domain of
the orthonormal tetrad is given by
\begin{equation}
g = \eta_{ab} e^{a} \otimes e^{b} \a; ,
\end{equation}
where
\begin{equation}
\bold{e}_{a}e_{b} = \eta_{ab} := \left\{ \begin{array}{rcl}
0 & \mbox{if} & a \ne b \\
1 & \mbox{if} & a = b = 1,2,3 \\
-1 & \mbox{if} & a = b = 4
\end{array} \right. \; .
\end{equation}
Hence,
\begin{equation}
\delta g = \eta_{ab} [(\delta e^{a}) \otimes e^{b}
+ e^{a} \otimes (\delta e^{b})] \; .
\end{equation}
However, we have seen in Eq.\ (\ref{result}) that
\begin{equation}
\delta e^{b} = - \lie{a}e^{b} + (\bold{a}\Gamma^{b}_{\; c})e^{c} \; .
\end{equation}
Because of the antisymmetry of the Ricci rotation matrix, the terms
in $\delta g$ that involve it cancel one another, and we end up with
the simple result
\begin{equation}
\delta g = - \lie{a} g \; .
\end{equation}
Thus, $\delta g = 0$ if and only if $\bold{a}$ is a Killing vector field.

Note should be taken of the fact that on the same manifold one can have
two isometric Riemannian spaces with {\em different} abstract metric
tensors.  For example, on a given Minkowski space, the type (b) generator
alters the abstract metric tensor $g$ if $\bold{a}$ is not equal to a
Killing vector, but the spacetime nevertheless remains flat.

\setcounter{equation}{0}
\section{Conclusions}
We believe that the study reported in this paper has brought us a little
closer to understanding the relationship between \gurses' work\cite{1,2}
and that of Torre and Anderson,\cite{3} who found that the only
generalized symmetries of the full Einstein equations, {\em in the
absence of Killing vectors}, are those associated with a constant scaling
of the spacetime metric and with what they call ``generalized
diffeomorphism symmetry.''  The latter symmetry appears to be equivalent
to \gurses' type (b) symmetry, and we have seen how the scaling symmetry
arises too.

While Torre and Anderson regard the generalized diffeomorphism symmetry
to be physically trivial, \gurses\ seems to suggest that {\em finite}
type (b) transformations may not be trivial.  What we know for certain
is that if one applies any finite type (b) transformation to any Petrov
type N spacetime, the Petrov type of the transformed spacetime is also
type N, and if one applies such a transformation to any Petrov type III
spacetime, the Petrov type of the transformed spacetime is also type
III.  We verified this by showing that the relevant Weyl conform tensor
invariant remains unchanged.  This result is consistent with the point
of view that finite type (b) transformations cannot be used to transform
one spacetime into another, but of course it is no proof of the
correctness of that view.  After we have had an opportunity to study
further the work of Torre and Anderson, we plan to return to this
question.

Finally, we have found that the symmetric traceless parts of the type (c)
and (\cbar) generators are subject to a differential constraint.  It is
possible that this constraint can only be satisfied if the initial
spacetime admits some Killing vectors, as, for example, in \gurses'
successful application of type (c) to a plane-wave metric to produce a
more general pp-wave.\cite{2}  We plan to investigate this possibility
in the immediate future.

It should be noted that all of the transformations considered in this
paper leave Minkowski space flat.  In particular, the type (c) and
(\cbar) transformations do not even alter the abstract metric tensor
$g$ of Minkowski space.  Therefore, in no way could these
transformations be used to generate stationary axisymmetric
spacetimes from Minkowski space, a key feature of the well-known
Kinnersley-Chitre transformations.

\setcounter{equation}{0}
\renewcommand{\theequation}{A.\arabic{equation}\roman{eqnpart}}
\section*{Appendix on Conventions and Notation}
{}From time to time we have included in our papers\cite{4} brief discussions
of our notations and conventions, but we have never attempted to bring
together all of the things that a reader might need in order to be able
to manipulate invariant geometrical objects efficiently.  This appendix
is our attempt to rectify this situation.  We anticipate that not only
the present paper but also our future publications concerning symmetries
of the Einstein equations will draw heavily upon the material in this
appendix.

We shall refer the reader to Hawking and Ellis\cite{5} for basic notions
of manifold theory, and merely provide our notations and conventions for
geometrical objects with which we shall assume he is familiar.

Unlike many other researchers, who seem content to deal only with
differential forms, we have found vector fields to be equally useful.
It has been our practice to introduce an {\em exterior algebra} over the
tangent space; i.e., $p$-vectors, where $p=0,1,2,3,4$.  We employ the
symbol $\bold{u} \wedge \bold{v}$ for the exterior product of $p$-vector
$\bold{u}$ and $q$-vector $\bold{v}$.  For two $p$-vectors of the same
degree, the inner product is defined in the obvious way, assuming that one
is dealing with a metric space.  However, we have found it useful to
introduce a more general Grassmann inner product,
$\bold{u} \stepdown \bold{v}$ that vanishes whenever the degree $p$ of
$\bold{u}$ is less than the degree $q$ of $\bold{v}$.  When $p \ge q$,
$\bold{u} \stepdown \bold{v}$ is that $p-q$-vector that has the property
\begin{equation}
(\bold{u} \stepdown \bold{v}) \cdot \bold{w} = \bold{u} \cdot
(\bold{v} \wedge \bold{w})
\end{equation}
for all $p-q$-vectors $\bold{w}$.
This Grassmann inner product enjoys the very useful property
\begin{equation}
(\bold{u} \stepdown \bold{v}) \stepdown \bold{w} =
\bold{u} \stepdown (\bold{v} \wedge \bold{w})
\label{grass1}
\end{equation}
for all $\bold{u},\bold{v},\bold{w}$, regardless of their degrees.
Moreover,
\begin{equation}
(\bold{u} \wedge \bold{v}) \stepdown \bold{w} =
(\bold{u} \stepdown \bold{w}) \bold{v} - \bold{u} (\bold{v} \stepdown
\bold{w})
\label{grass2}
\end{equation}
for all $1$-vectors $\bold{u},\bold{w}$ and $p$-vectors $\bold{v}$.

Similarly we introduce an {\em exterior algebra} over the cotangent space,
i.e., the dual space of differential forms.  Between differential forms we
suppress the wedge symbol of exterior multiplication.  For two $p$-forms
of the same degree, the inner product is defined in the obvious way.
More generally we introduce another Grassmann inner product, $u \stepup v$
that vanishes whenever the degree $p$ of $u$ is greater than the degree
$q$ of $v$.  When $p \le q$, $u \stepup v$ is that $q-p$-form that has the
property
\begin{equation}
w \cdot (u \stepup v) = (w u) \stepup v
\end{equation}
for all $q-v$-forms
$w$.  This Grassmann inner product satisfies
\begin{equation}
w \stepup (v \stepup u) = (w v) \stepup u
\label{grass3}
\end{equation}
for all $u,v,w$, regardless of their degrees, and
\begin{equation}
w \stepup (v u) = v (w \stepup u) - (w \stepup v) u
\label{grass4}
\end{equation}
for all $1$-forms $u,w$ and $p$-forms $v$.
Neither of these Grassmann inner products should be confused with the
{\em contraction} of a vector field and a differential form, which is defined
even in the absence of a metric.  The contraction of the $q$-vector field
$\bold{a}$ and the $p$-form $b$ we shall denote by $\bold{a} b$, with the
form written to the {\em right} of the vector field.  It is simply the
result of applying $b$ as a linear functional to $\bold{a}$.  On the
other hand, if we write the $p$-form $b$ to the {\em left} of the $q$-vector
field $\bold{a}$, that object, $b\bold{a}$, will be a \tensor{p}{q}.
The most frequently used of these will be \tensor{1}{1}s and \tensor{2}{2}s.

Corresponding to any given chart $\{ x^{b} (b=1,\cdots,4) \}$ we can
introduce a {\em natural basis} $\{ \bold{\partial}_{\alpha} \;
(\alpha=1,\cdots,4) \}$ for the tangent space, and write, for example,
$\bold{a}=a^{\alpha} \bold{\partial}_{\alpha}$.  Similarly, $\{
dx^{\beta} \; (\beta = 1,\cdots,4) \}$, a natural basis for the cotangent
space, is defined by requiring
\begin{equation}
\bold{\partial}_{\alpha} dx^{\beta} = \delta_{\alpha}^{\; \beta},
\end{equation}
where $\delta_{\alpha}^{\; \beta}$ is the
Kronecker delta.  An {\em orthonormal} basis for the tangent space
will be denoted by $\{ \bold{e}_{a} \; (a=1,\cdots,4) \}$.  The
corresponding orthonormal basis for the cotangent space will be
denoted by $\{ e^{b} \; (b=1,\cdots,4) \}$, where
\begin{equation}
\bold{e}_{a} e^{b} = \delta_{a}^{\; b}.
\end{equation}
A {\em null tetrad} basis for the tangent
space will be denoted by $\{ \bold{k},\bold{m},\bold{t},\bold{t}^{*}
\}$, where the nonvanishing inner products are
\begin{equation}
\bold{k} \cdot \bold{m} = \bold{t} \cdot \bold{t}^{*} = +1.
\end{equation}
The corresponding basis for the cotangent space is $\{ k,m,t,t^{*} \}$,
where $\bold{k}m = \bold{m}k = \bold{t} t^{*} = \bold{t}^{*} t = +1$, and
all other contractions vanish.  These are notations and conventions of
R.\ Sachs\cite{6} that we adopted long ago.

If $u$ is any $p$-form, then $\star u$ is the $4-p$-form
\begin{equation}
\star u := -(-1)^{p} u \stepup (e^{1}e^{2}e^{3}e^{4}).
\end{equation}
Moreover,
\begin{equation}
\star \star u = -(-1)^{p} u.
\end{equation}
In particular, for any orthonormal basis,
\begin{subequations}
\begin{eqnarray}
\star 1 & = & - e^{1} e^{2} e^{3} e^{4} \; , \\
\star e^{4} & = & - e^{1} e^{2} e^{3} \; , \\
\star e^{3} & = & - e^{4} e^{1} e^{2} \; , \\
\star (e^{1} e^{2}) & = & e^{4} e^{3} \; ,
\end{eqnarray}
\end{subequations}
and other valid relations can be obtained by cyclic permutations of the
indices $1,2,3$.  Similarly, a duality operator for the tangent space
can be defined by
\begin{equation}
\star \bold{u} := - (\bold{e}^{1} \wedge \bold{e}^{2} \wedge
\bold{e}^{3} \wedge \bold{e}^{4}) \stepdown \bold{u}.
\end{equation}

In the case of $2$-forms we often express the $\star$ operator as a
\tensor{2}{2}
\begin{equation}
\striated{D} := \frac{1}{4} e^{a} e^{b} \epsilon_{abcd}
\bold{e}^{c} \wedge \bold{e}^{d} .
\label{Ddef}
\end{equation}
This duality operator $\striated{D}$ has the property
\begin{equation}
\striated{D}^{2} = - \striated{I}^{(2)},
\end{equation}
where
\begin{subequations}
\begin{eqnarray}
\striated{I}^{(2)} & := & \frac{1}{2} \striated{I} \wedge \striated{I}
\label{I2def}
\; , \\
\striated{I} & = & e^{b} \bold{e}_{b} \; .
\label{Idef}
\end{eqnarray}
\end{subequations}
Hence, the eigenvalues of $\striated{D}$ are $\pm i$.
The projection operator that projects onto self-dual $2$-forms is
\begin{equation}
\striated{P} := \frac{1}{2} (\striated{I}^{(2)} - i \striated{D}) \; .
\label{Pdef}
\end{equation}

The $2$-forms that correspond to eigenvalue $+i$ will be called
{\em self-dual}.  A convenient basis for self-dual $2$-forms is
\begin{equation}
B_{1} := kt \; , \; B_{0} := km+tt^{*} \; , \; B_{-1} := mt^{*} \; .
\label{Bdef}
\end{equation}
Two extremely useful identities are
\begin{subequations}
\begin{eqnarray}
\bold{B}_{I} \stepdown (\bold{B}_{J}^{*} \stepdown \striated{I})
& = & \bold{B}_{J}^{*} \stepdown (\bold{B}_{I} \stepdown \striated{I})
\label{id3} \; , \\
B_{I} (\striated{I} \stepup B_{J}^{*})
& = & - B_{J}^{*} (\striated{I} \stepup B_{I}) \; .
\label{id4}
\end{eqnarray}
\end{subequations}

The connection $1$-forms are defined by
\begin{equation}
\Gamma_{ab} := d\bold{e}_{a} \stepdown \bold{e}_{b}
\end{equation}
in the case of an orthonormal tetrad.  This so-called {\em Ricci
rotation matrix} is antisymmetric.  We then have
\begin{subequations}
\begin{eqnarray}
d\bold{e}_{a} & = & \Gamma_{a}^{\; b} \bold{e}_{b} \; , \\
de^{b} & = & e^{a} \Gamma_{a}^{\; b} \; .
\end{eqnarray}
\end{subequations}
If $\bold{a}$ is any $1$-vector field, then
\begin{equation}
d\bold{a} = d(a^{c}\bold{e}_{c}) = da^{c} \bold{e}_{c} + a^{c}
d\bold{e}_{c} = e^{b} (\nabla_{b} a^{c}) \bold{e}_{c}
\end{equation}
is the covariant derivative.

In the case of a null tetrad we introduce the complex connection
$1$-forms
\begin{equation}
u := P + i Q \; , \; P := d\bold{k} \stepdown \bold{m} \; , \;
iQ := d\bold{t} \stepdown \bold{t}^{*} \; , \;
v := d\bold{k} \stepdown \bold{t} \; , \;
w := d\bold{m} \stepdown \bold{t}^{*} \; .
\label{uvwdef}
\end{equation}
Then
\begin{subequations}
\begin{eqnarray}
d\bold{k} & = & P \bold{k} + v^{*} \bold{t} + v \bold{t}^{*} \; , \\
d\bold{m} & = & - P \bold{m} + w \bold{t} + w^{*} \bold{t}^{*} \; , \\
d\bold{t} & = & - w^{*} \bold{k} - v \bold{m} + iQ \bold{t} \; ,
\end{eqnarray}
\end{subequations}
and
\begin{subequations}
\begin{eqnarray}
dk & = & P k + v^{*} t + v t^{*} \; , \\
dm & = & - P m + w t + w^{*} t^{*} \; , \\
dt & = & - w^{*} k - v m + iQ t \; .
\end{eqnarray}
\end{subequations}
The latter relations are equivalent to the following ones:
\begin{subequations}
\begin{eqnarray}
dB_{1} & = & B_{1} u - B_{0} v \; , \\
dB_{0} & = & 2 (B_{1} w - B_{-1} v) \; , \\
dB_{-1} & = & B_{0} w - B_{-1} u \; .
\end{eqnarray}
\end{subequations}

The curvature tensor
\begin{equation}
\striated{R} = \frac{1}{4} e^{a} e^{b} R_{abcd} \bold{e}^{c} \wedge
\bold{e}^{d}
\end{equation}
is, like $\striated{D}$, a \tensor{2}{2};
namely,
\begin{equation}
\striated{R} := \frac{1}{2} \bold{e}^{a} \wedge d^{2}\bold{e}_{a} \; .
\label{Rdef}
\end{equation}
For any \tensor{p}{1}\ $\bold{\alpha}$, one has
\begin{equation}
d^{2}\bold{\alpha} = \striated{R} \stepdown \bold{\alpha}.
\label{d2}
\end{equation}
In this formalism the Bianchi integrability condition and the
algebraic curvature identity assume the simple forms
\begin{subequations}
\begin{eqnarray}
d\striated{R} & = & 0 \; , \label{Bianchi} \\
\striated{R} \stepdown \striated{I} & = & 0 \; ,
\label{algebraic}
\end{eqnarray}
\end{subequations}
respectively.

The duality operator $\striated{D}$ can be used to separate $\striated{R}$ into
its Ricci and Weyl parts.  Thus,
\begin{equation}
\striated{R} = \striated{C} + \striated{E} + \frac{R}{12}
\striated{I}^{(2)} \; ,
\end{equation}
where
\begin{subequations}
\begin{eqnarray}
\striated{E} & := & \striated{P} \striated{R} \striated{P}^{*}
  + \striated{P}^{*} \striated{R} \striated{P} \; , \\
\label{Edef}
\striated{C} & := & \striated{P} \striated{R} \striated{P}
  + \striated{P}^{*} \striated{R} \striated{P}^{*}
  - \frac{R}{12} \striated{I}^{(2)} \; .
\label{Cdef}
\end{eqnarray}
\end{subequations}
In general the relation between the Riemann tensor and the connection
$1$-forms is given by
\begin{equation}
\frac{1}{2} e^{a} e^{b} R_{abc}^{\rule{1.0em}{0.0em} d} =
d\Gamma_{c}^{\; d} - \Gamma_{c}^{\; r} \Gamma_{r}^{\; d}
\end{equation}
if an orthonormal tetrad is used.  If one uses a  null tetrad these
equations can be expressed in the form
\begin{subequations}
\begin{eqnarray}
dv-uv & = & c_{2} B_{-1} + c_{1} B_{0} + (c_{0}+\frac{R}{12}) B_{1}
 \label{dv}
 \nonumber \\
& & + \frac{1}{2} S_{kk} B_{1}^{*} + \frac{1}{2} S_{kt} B_{0}^{*}
 + \frac{1}{2} S_{tt} B_{1}^{*} \; , \\
du+2vw & = & - 2[c_{1} B_{-1} + (c_{0}-\frac{R}{24}) B_{0} + c_{-1} B_{1}
 \label{du}
 \nonumber \\
& & + \frac{1}{2} S_{kt^{*}} B_{-1}^{*} + \frac{1}{2} S_{tt*} B_{0}^{*}
 - \frac{1}{2} S_{mt} B_{1}^{*}] \; , \\
dw+uw & = & (c_{0}+\frac{R}{12}) B_{-1} + c_{-1} B_{0} + c_{-2} B_{1}
 \label{dw}
 \nonumber \\
& & + \frac{1}{2} S_{t^{*}t^{*}} B_{-1}^{*} - \frac{1}{2} S_{mt^{*}} B_{0}^{*}
 + \frac{1}{2} S_{mm} B_{1}^{*} \; ,
\end{eqnarray}
\end{subequations}
where $S_{ij}$ is the traceless part of the Ricci tensor.

\section*{Acknowledgement}
This work was supported in part by grants PHY-91-16681 and PHY-92-08241
from the National Science Foundation.  We also thank M.\ \gurses,
C.\ Torre and I.\ Anderson for sending us preprints of the Physical
Review Letters articles in which their exciting discoveries were
announced and for various prepublication discussions that we had with
these authors.

\newpage
\section*{References:}
\begin{enumerate}
\item
M.\ \gurses, Phys.\ Rev.\ Lett.\ {\bf 70}, 367 (1993).
\item
M.\ \gurses, Phys.\ Rev.\ Lett.\ {\bf xx}, xxx (1993).
In this paper \gurses\ generalized his original definition of the type
(c) symmetry, including an additional term $s$ in $(\delta \Gamma)\sigma$.
It should be noted that the one known example of a nontrivial application
of the type (c) or (\cbar) generator that is reported in this paper can
be carried out entirely within the $\striated{S}=0$ or $s=0$ framework of
our Theorem 2.
\item
C.\ Torre and I.\ Anderson, Phys.\ Rev.\ Lett. {\bf xx}, xxx (1993).
\item
Our formalism has been described in various places.  Some useful
references are:  F.\ J.\ Ernst, J.\ Math.\ Phys.\ {\bf 15}, 1409 (1974);
F.\ J.\ Ernst, J.\ Math.\ Phys.\ {\bf 19}, 489 (1978); I.\ Hauser and
F.\ J.\ Ernst, J.\ Math.\ Phys.\ {\bf 19}, 1316 (1978).
\item
S.\ W.\ Hawking and G.\ F.\ R.\ Ellis, {\em The Large Scale Structure of
Space-Time}, Cambridge University Press, 1973.
\item
Our notation ultimately derives from an article by R.\ K.\ Sachs in {\em
Relativity, Groups and Topology, Les Houches 1963}, ed.\ C.\ DeWitt and
B.\ DeWitt, Gordon and Breach Science Publishers (1964), pp. 521--562.
\end{enumerate}

\end{document}